%
% the following is to use blackboard bold fonts --
\let\useblackboard=\iftrue
%
% activate this if you don't have them.
%\let\useblackboard=\iffalse
%
% You might also need to remove this line.
\newfam\black
\input harvmac.tex
\def\Title#1#2{\rightline{#1}
\ifx\answ\bigans\nopagenumbers\pageno0\vskip1in%
\baselineskip 15pt plus 1pt minus 1pt
\else%\special{papersize=11in,8.5in}%
\def\listrefs{\footatend\vskip 1in\immediate\closeout\rfile\writestoppt
\baselineskip=14pt\centerline{{\bf References}}\bigskip{\frenchspacing%
\parindent=20pt\escapechar=` \input
refs.tmp\vfill\eject}\nonfrenchspacing}
\pageno1\vskip.8in\fi \centerline{\titlefont #2}\vskip .5in}

\ifx\answ\bigans\def\tcbreak#1{}\else\def\tcbreak#1{\cr&{#1}}\fi
\useblackboard
\message{If you do not have msbm (blackboard bold) fonts,}
\message{change the option at the top of the tex file.}
\font\blackboard=msbm10 scaled \magstep1
\font\blackboards=msbm7
\font\blackboardss=msbm5
%\newfam\black
\textfont\black=\blackboard
\scriptfont\black=\blackboards
\scriptscriptfont\black=\blackboardss

\else

\fi
% *************************************
%\draft
%
\def\yboxit#1#2{\vbox{\hrule height #1 \hbox{\vrule width #1
\vbox{#2}\vrule width #1 }\hrule height #1 }}
\def\fillbox#1{\hbox to #1{\vbox to #1{\vfil}\hfil}}
\def\ybox{{\lower 1.3pt \yboxit{0.4pt}{\fillbox{8pt}}\hskip-0.2pt}}

\def\comments#1{}
\def\cc{{\rm c.c.}}

\def\p{\partial}

\def\eps{\epsilon}

\def\CD{{\cal D}}

\def\CN{{\cal N}}

\def\CL{{\cal L}}

\def\nl{\hfill\break}

\def\I{I}

\def\II{\relax{I\kern-.10em I}}
\def\IIa{{\II}a}
\def\IIb{{\II}b}

\def\hk{hyperk\"ahler\  }

\def\IZ{\relax\ifmmode\mathchoice
{\hbox{\cmss Z\kern-.4em Z}}{\hbox{\cmss Z\kern-.4em Z}}
{\lower.9pt\hbox{\cmsss Z\kern-.4em Z}}
{\lower1.2pt\hbox{\cmsss Z\kern-.4em Z}}\else{\cmss Z\kern-.4em
Z}\fi}
\def\IB{\relax{\rm I\kern-.18em B}}
\def\IC{{\relax\hbox{$\inbar\kern-.3em{\rm C}$}}}
\def\ID{\relax{\rm I\kern-.18em D}}
\def\IE{\relax{\rm I\kern-.18em E}}
\def\IF{\relax{\rm I\kern-.18em F}}
\def\IG{\relax\hbox{$\inbar\kern-.3em{\rm G}$}}
\def\IGa{\relax\hbox{${\rm I}\kern-.18em\Gamma$}}
\def\IH{\relax{\rm I\kern-.18em H}}
\def\II{\relax{\rm I\kern-.18em I}}
\def\IK{\relax{\rm I\kern-.18em K}}
\def\IP{\relax{\rm I\kern-.18em P}}
%\def\IX{\relax{\rm X\kern-.01em X}}
%this doesn't work

\def\inbar{\,\vrule height1.5ex width.4pt depth0pt}

\def\p{\partial}

\font\cmss=cmss10 \font\cmsss=cmss10 at 7pt
\def\IR{\relax{\rm I\kern-.18em R}}

\def\ker{{\rm ker\ }}
\def\cok{{\rm cok\ }}

%%
%%% END MACROS FROM GREG
%%
%%  MERGING  MIKE AND GREG'S BLACKBOARD FONTS
\def\BR{\IR}

\def\BR{\IR}

\Title{ \vbox{\baselineskip12pt\hbox{hep-th/9604198}
\hbox{RU-96-24}}}
{\vbox{
\centerline{Gauge Fields and D-branes}}}
\centerline{Michael R. Douglas}
\smallskip
\smallskip
\centerline{Department of Physics and Astronomy}
\centerline{Rutgers University }
\centerline{Piscataway, NJ 08855-0849}
\centerline{\tt mrd@physics.rutgers.edu}
\bigskip
%\centerline{others}
%\smallskip
%\centerline{xxx}
%\centerline{\tt xxx}
\bigskip
\noindent
We prove that self-dual gauge fields in type \I\ superstring theory are
equivalent to configurations of Dirichlet $5$-branes,
by showing that the world-sheet theory of a Dirichlet $1$-brane moving
in a background of $5$-branes includes an ``ADHM sigma model.''
This provides an explicit construction of the equivalent
self-dual gauge field.  We also discuss type \II.

\Date{April 1996}
%\draft
%
\lref\dlp{J.~Dai, R.~G.~Leigh and J.~Polchinski, Mod. Phys. Lett. {\bf A4}
(1989) 2073.}
\lref\pol{J.~Polchinski, Phys.~Rev.~Lett.~75 (1995) 4724-4727;
hep-th/9510017.}
\lref\chs{C. Callan, J. Harvey and A. Strominger, Nucl. Phys.
{\bf B367} (1991) 60.}
\lref\witten{E. Witten, ``Small Instantons in String Theory,'' hep-th/9511030.}
\lref\adhm{M. F. Atiyah, V. G. Drinfeld, N. J. Hitchin, and Y. I. Manin,
Phys. Lett. {\bf A65} (1978) 185.}
\lref\witadhm{E. Witten, ``Sigma Models and the ADHM Construction of
Instantons,'' J. Geom. Phys. 15 (1995) 215-226.}
\lref{\witstr}{E. Witten, ``Some Comments On String Dynamics,''
hep-th/9507121.}
\lref\kn{P.B. Kronheimer and H. Nakajima, ``Yang-Mills instantons
on ALE gravitational instantons,'' Math. Ann. {\bf 288} (1990) 263.}
\lref\gimon{E. G. Gimon and J. Polchinski,
``Consistency Conditions for Orientifolds and D-manifolds,'' hep-th/9601038.}
\lref\hst{P. Howe, G.Sierra and P. Townsend, Nucl. Phys. {\bf B221} (1983)
331;\nl
G. Sierra and P. Townsend,
``The gauge invariant N=2 supersymmetric
sigma model with general scalar potential,''  Nucl. Phys. B233 (1984) 289.}
\lref\douglas{M.~R.~Douglas, ``Branes within Branes,'' hep-th/9512077.}
\lref\dm{M. Douglas and G. Moore, ``D-Branes, Quivers, and ALE Instantons,''
hep-th/9603167.}
\nref\polwit{J.~Polchinski and E.~Witten, ``Evidence for Heterotic-Type I
String Duality,'' hep-th/9510169.}
\lref\joerev{See S. Chaudhuri, C. Johnson, and J. Polchinski,
``Notes on D-Branes,'' hep-th/9602052, for a recent review.}
\lref\kn{P.B. Kronheimer and H. Nakajima, ``Yang-Mills instantons
on ALE gravitational instantons,'' Math. Ann. {\bf 288} (1990) 263.}
\lref\etal{M. Douglas, G. Moore, N. Nekrasov, and
S. Shatashvili, ``D-branes and the construction of
$SO(w)$ instantons on ALE spaces,'' to appear}
%
% forward equation references
%
\newsec{Introduction}

Recent work has demonstrated the power of the Dirichlet brane
\refs{\dlp,\pol}\ in constructing highly non-trivial solutions
of superstring theory \joerev.

Although at first glance these solutions look very different from those
in which a conventional metric and gauge fields
satisfy conventional equations of motion (with stringy corrections),
it is known that in many cases they are to be regarded as extreme
limits of field theory solutions, singular in field theory
but non-singular in string theory.  This stringy resolution of the
singularity can already be seen in the low energy limit, as due
to additional massless degrees of freedom in the effective field theory.

The example currently best understood is the Yang-Mills instanton.
In type \I\
superstring theory this becomes a $5$-brane with the appropriate charge
and action to be a Dirichlet brane.
Witten argued that the D$5$-brane
world-volume is identical to the zero scale size limit of the instanton
\witten, by showing that
the space of vacua of a system of parallel
D$5$-branes is isomorphic to the
moduli space of instantons on $\BR^4$.
Indeed, the physical definition of these vacua (solutions
of D-flatness conditions of a supersymmetric gauge theory) is identical
to the ADHM construction \adhm,
which produces instanton moduli space as a \hk\ quotient.

The ADHM construction also provides an explicit construction of
the actual gauge potential for these instantons.
To reproduce this physically,
we should insert a very heavy charged probe, point-like in the
four dimensions transverse to the $5$-branes whose position is
well defined.  Its motion will then depend locally on the vector potential
defined at a point.

The natural probe is another D-brane.  A D$1$-brane extending parallel
to the $5$-branes will be pointlike in the four dimensions containing the
instanton, and its world-sheet
action will be a sigma model describing propagation
of a heterotic string soliton in this gauge field.
It has $(0,4)$ supersymmetry and its field content was described in \douglas.
The $5$ and $9$-brane fields will play the role of couplings.
Since these are ADHM data, we might expect the world-sheet theory to
be equivalent to Witten's ADHM sigma model \witadhm.
We now show that this is true.

\newsec{Type \I}

Consider a $1$-brane moving
in the presence of $k$ $5$-branes and $N$ $9$-branes in type \I\ theory.
We will conform to the notations of \refs{\witadhm,\dm,\etal} where 
possible.\footnote*{
Notations we changed: $X$ in \witadhm\ is now $b$; $Y$ in \douglas\ is
now $\phi$.}

We take $1$-branes in the $01$-plane, and $5$-branes in $012345$.
The Lorentz symmetry becomes $SO(1,1)\times SO(4)_I \times SO(4)_E$.
We index doublets of the two $SU(2)$'s in $SO(4)_E$ as $A$ and $Y$,
and doublets of the two $SU(2)$'s in $SO(4)_I$ as $A'$ and $\tilde A'$.
All of these indices will be raised and lowered as
$v_A = \epsilon_{AB} v^B$.
The $(0,4)$ supersymmetries are parameterized by $\eta^{AA'}_+$.
We index $5$-branes with $m$ and $9$-branes with $M$.
The indices $m$ will be raised and lowered with the $Sp(2k)$ invariant
antisymmetric tensor $\eps^{mn}$.
For consistency with \witadhm,
let the index $a$ be the union of choices from $M$ and $Y\times m$,
and let $Y'$ be an alternate name for $m$.

The $5$-brane fields include $Sp(2k)$ gauge theory,
hypermultiplets in the antisymmetric tensor representation,
and ``half-hypermultiplets''
in the $(2k,N)$ of $Sp(2k)\times SO(N)$.
Their scalar components are
$X^{AY}=\eps^{AB}\eps^{YZ}X^*_{BZ}$
and $h_{M}^{Am}=i\eps^{mn}\eps_{AB}(h^*)_{Bn}^{M}$.
These will become couplings in the $1$-brane world-sheet theory.

The $(0,4)$ multiplets come in two varieties,
`standard' with $\delta b^{Ai} = \eta^A_{A'} \psi^{A'i}$,
and `twisted' with $\delta \phi^{A'j} = \eta_A^{A'} \chi^{Aj}$.

The world-sheet fields are then
$$\matrix{
\hbox{boson}&\hbox{fermion}&\hbox{sector}&\cr
&&&&\cr
b^{AY}&\psi^{A'Y}_-&DD&\hbox{positions transverse to $5$-brane}\cr
b^{A'\tilde A'}&\psi^{A\tilde A'}_-&DD&\hbox{positions within $5$-brane}\cr
\phi^{A'm}&\chi^{A m}_-&DN5&\hbox{$1-5$ strings }\cr
&\chi^{Y m}_+&DN5&
	\hbox{GSO correlates $SO(1,1)$ and $SO(4)_E$ chirality}\cr
&\lambda^{M}_+&DN9&
	\hbox{GSO selects $SO(1,1)$ chirality}\cr
&\lambda^a_+&&\hbox{An alternate notation for the set
	$(\chi^{Ym}_+,\lambda^M_+)$.}
}$$
Here
$b^{AY}=\eps^{AB}\eps^{YZ}\bar b_{BZ}$, and
$\phi^{A'm}=\eps^{A'B'}\eps^{mn}\bar \phi^{B'n}$.
The $\lambda$'s are real.

It is useful to first consider
subsectors involving a subset of the branes.  We will use the fact that
upon combining subsectors, the action is the sum of the action for each
subsector, plus possible new couplings involving more than one subsector.
The $19$ subsector of $b$, $\psi$ and $\lambda$
preserves $(0,8)$ supersymmetry and is just the heterotic soliton theory
of \polwit\ with a free action.

The $15$ subsector
can be derived by dimensional reduction of a $d=6$, $\CN=1$
$59$ theory in the $016789$ plane.  Due to the $\Omega$ projection this is
an ``$SO(1)$ gauge theory,'' but nevertheless the couplings to $b^{AY}$ are
determined.
They give the DN modes $\phi$ and $\chi$
a mass proportional to the distance between the $1$ and $5$-branes.
Translation symmetry in the $6789$ plane allows only the combination
$(X^{AY}_{m\bar n} - b^{AY}\delta_{m\bar n})$ to appear:
\eqn\onefive{\eqalign{
\CL = \CL_{free} + &\phi_{A' m} \phi^{A'n}
	(X^{AY}_{mn} - b^{AY}\delta_{mn})^2 +\cr
	&\chi^{A m}_- \chi^{Yn}_+
		(X^{AY}_{m n} - b^{AY}\delta_{m n}) +
	\psi^{A'Y}_- \chi_{Y+}^{m} \phi_{A' m}
		 + {\rm c.c.}
}}
(and couplings to the fermionic partners of $X$).
The last of these is required by $(0,4)$ supersymmetry.

The $159$ couplings respect only $(0,4)$ supersymmetry.
These can be computed by world-sheet techniques; for example
\eqn\onefivenine{
\CL = \lambda^{M}_+ \chi^{A}_{m-} h_{AM}^{m}
}
comes from a correlator of three twist fields, which is evidently non-zero.
%A $h^2\phi^2$ potential could also be present.

By now imposing $(0,4)$ supersymmetry, we have enough information to get
the complete Lagrangian.  In \witadhm\ the general $(0,4)$ Lagrangian with
this field content was worked out, and is given in equation (2.21):
\eqn\finter{\eqalign{
L = L_{free} &-{i\over 4}\int d^2\sigma\ \lambda_+^a \left(
	\eps^{BD}{\p C^a_{BB'}\over \p b^{DY}} \psi_-^{B'Y} +
	\eps^{B'D'}{\p C^a_{BB'}\over \p \phi^{D'Y'}} \chi_-^{BY'} \right)\cr
&+ {1\over 8}\int d^2\sigma\ \eps^{AB}\eps^{A'B'} C^a_{AA'}C^a_{BB'}.
}}
The functions $C$ determine the supersymmetry transformations
of the left moving fermions
\eqn\leftsusy{
\delta\lambda^{a}_+ = \eta^{AA'}_+ C^a_{AA'}
}
and are determined in terms of four constant matrices $M$, $N$, $D$ and $E$:
\eqn\csoln{
C^a_{AA'} = M^a_{AA'} + b_{AY} N^{aY}_{A'}
+ \phi_{A'}^{Y'} D^a_{AY'} + \phi_{A'}^{Y'} b_{AY} E^a_{YY'}.
}
The condition for $(0,4)$ supersymmetry is then
\eqn\zerofour{
0 = \sum_a C^a_{AA'}C^a_{BB'} + C^a_{BA'}C^a_{AB'}.
}
Comparing with \onefive\ determines
\eqn\detonefive{
C^{Ym}_{AA'} = \phi_{A'n} (X^{AY}_{mn} - b^{AY}\delta_{mn}).
}
The coupling \onefivenine\ will be reproduced by
\eqn\detonefivenine{
C^{M}_{AA'} = h^M_{Am} \phi_{A'm}
}
which also implies the existence of an $h^2\phi^2$ term in the potential.

This completes the derivation of the Lagrangian, and indeed the couplings
of the sigma model are given by ADHM data.
Thus the $1$-brane world-volume theory is essentially
Witten's ADHM sigma model
(with the additional free fields $b^{A'\tilde A'}$ and $\psi^{A\tilde A'}_-$).
The condition for $(0,4)$ world-sheet supersymmetry \zerofour\ is
\eqn\zfour{
0 =
h^{M(A}_{m}h^{B)}_{Mn} + \eps^{pq}\eps_{YZ} X^{(A~Y}_{mp} X^{B)Z}_{nq}
}
which is just the D-flatness condition for the $5$-brane theory.
As is often the case, the conditions for extended supersymmetry in the sigma
model are the equations of motion in the target space.

\subsec{The ADHM connection}

Integrating out massive degrees of freedom on the $1$-brane
will produce a conventional gauged sigma model.
This was done explicitly in \witadhm,
to produce the self-dual gauge connection
corresponding to the ADHM data $X$ and $h$.

For generic ADHM data $X\ne 0$,
the potential $V=\sum C^2\sim \phi^2(X^2+b^2)$ gives mass to $\phi$,
leaving the fields $b$ to describe the location of the $1$-brane in
four dimensions.
The Yukawa couplings of \finter, \detonefive\ and \detonefivenine\ become,
taking $\phi=0$, and suppressing the $mnMN$ indices,
\eqn\yuk{
\left(\matrix{\chi_{1}&\chi_{2}}\right)_-.
\left(\matrix{X^{11}-b^{11}&X^{12}-b^{12}&h_{M}^{1m}\cr
	X^{21}-b^{21} &X^{22}-b^{22} &(h^{2~tr})_m^{M}}\right)
\left(\matrix{\chi^{1}\cr\chi^{2}\cr\lambda}\right)_+
}
These couplings will pair $\chi_-$ with $4k$ fermions from the set
$\lambda^a_+$, leaving an $N$-dimensional subspace of massless modes
varying with $b$.  This bundle of massless modes is the resulting gauge
bundle, and projecting the original flat connection onto this subbundle
will produce the ADHM connection.

Referring to the operator in \yuk\ as $\CD^+$,
the subbundle is $E\equiv {\rm ker~} \CD^+$.
Given an explicit orthonormal
basis $v^a_{i}(b)\in E$, the connection can be written
by changing variables to massless fermions $\lambda_{i+}$:
\eqn\mlf{
\lambda^a_+ = \sum_{i=1}^N v^a_{i}(b)\lambda_{i+}.
}
Inserting this in the kinetic term $\lambda^a_+\p_-\lambda^a_+$
produces the Lagrangian
\eqn\gaugeact{\eqalign{
\CL &=
 \lambda_{i+} (v^a_{i}(b))^{-1}\p_-v^a_{j}(b)\lambda_{j+} \cr
 &=
 \lambda_{i+} (\delta^{ij}\p_- + A^{ij}_{AY}
 	(b^{AY}) \p_- b^{AY})\lambda_{j+}
}}
with $A^{ij}_{AY}(b) = (v^a_{i})^{-1}\p v^a_{j}/\p b^{AY}$.

A last point to mention is that there can also be
a direct coupling to the gauge field on the $9$-branes,
\eqn\direct{
\int d^2\sigma\ A^{(b)~MN}_\mu(b)\p_- b^\mu \lambda^M_+\lambda^N_+ .
}
Including this coupling,
the conditions for $(0,4)$ supersymmetry should be (more or less)
that the sum of the bare gauge field $A^{(b)}$ and the ADHM gauge field
is self-dual.
Physically this is a necessary addition (e.g. in Minkowski space we could
have other gluons propagating) but for purposes of constructing self-dual
gauge fields $A^{(b)}$ can be set to zero.
In terms of the 2d field theory, in Euclidean
four dimensions these couplings are redundant.

\newsec{Type \II}

Repeating the construction for a type \II\ theory will produce $U(N)$
self-dual connections.
The most direct analog of the type \I\ construction is to take
$8$, $4$ and $0$-branes in type \IIa\ theory.
The world-volume theories are similar except that we do not impose the
world-sheet orientation projection $\Omega=1$, and we reduce the
dimension on each brane by one.

It is convenient to first derive the (unphysical)
theory of $9$, $5$ and $1$ branes in
type \IIb\ and then do dimensional reduction.
The basic $1$-brane now has $(8,8)$ supersymmetry and there are
right movers $\psi_+$ and a $U(1)$ gauge field.
The gauge fermions $\lambda$ are still chiral (this was due to GSO)
but are now complex, as are the fields $\phi$ and $\chi$
and the couplings $h$.
We distinguish fundamental and antifundamental indices $mM$ and $\bar m\bar M$.
The full list of fields is
$$\matrix{
\hbox{boson}&\hbox{fermion}&\hbox{sector}&\cr
&&&\cr
b^{AY}&\psi^{A'Y}_-&DD&\cr
b^{A'\tilde A'}&\psi^{A\tilde A'}_-&DD&\cr
A_{--},A_{++}&\psi^{AA'}_+,\psi^{\tilde A' Y}_+&DD&
	\hbox{$U(1)$ gauge field and $(0,8)$ gauginos}\cr
\phi^{A'\bar m}&\chi^{A\bar m}_-&DN5&\hbox{$1-5$ strings (complex fields)}\cr
\bar\phi^{A'm}&\bar\chi^{Am}_-&DN5&\hbox{charge conjugates}\cr
&\chi^{Y\bar m}_+,\bar\chi^{Ym}_+&DN5&\cr
&\lambda^{\bar M}_+,\bar\lambda^M_+&DN9&	\cr
}$$

The 15 sector now has $(4,4)$ world-sheet supersymmetry.
The interactions include those in the gauged kinetic term,
the $U(1)$ D-term and the gaugino couplings,
and those following from the $(2,2)$-superspace superpotential
$W=(X^{AY}_{m n} - b^{AY}\delta_{mn}) \phi_{A' m} \phi^{A'n}$:
\eqn\IIonefive{\eqalign{
\CL = \CL_{kin} &+ 	|\phi_{1 m}|^2 - |\phi_{2 m}|^2
		+	(\psi^{A'Y}_- \bar\chi_{Y+}^{m} +
				\psi^{AA'}_+ \bar\chi_{A-}^{m})
			\phi_{A'\bar m}  + {\rm c.c.}\cr
		&+(\phi_{A' m} \bar\phi_{A'\bar m})^2
		+\phi_{A'\bar m} \bar\phi^{A'n}
	(X^{AY}_{m\bar n} - b^{AY}\delta_{m\bar n})^2 +\cr
	&\chi^{A\bar m}_- \bar\chi^{Yn}_+
		(X^{AY}_{m\bar n} - b^{AY}\delta_{m\bar n}) .
}}
The extra $\phi^4$ potential terms have no effect on the branch $\phi=0$.

The 159 couplings are those following from a complex
version of \detonefivenine,
\eqn\IIdetonefivenine{\eqalign{
C^{M}_{AA'} &= \bar h^{\bar M}_{A m} \phi_{A'\bar m} \cr
\bar C^{\bar M}_{AA'} &= h^{M}_{A\bar m} \bar \phi_{A' m}
}}
leading to the $\lambda$ Lagrangian
\eqn\IIonefivenine{\eqalign{
\CL = \int d^2\sigma\ &\bar\lambda^{M}_+
		(\p_{--}+iA_{--})\lambda^{\bar M}_+ +
\lambda^{\bar M}_+ \bar\chi^{A}_{ m-} h_{A M}^{\bar m} + \cc \cr
&+ {1\over 4}\int d^2\sigma\ \eps^{AB}\eps^{A'B'} C^M_{AA'}C^{\bar M}_{BB'}.
}}

To verify $(0,4)$ supersymmetry, we
follow the analysis in \witadhm, promoting $\lambda$ and $C$ to
complex fields.
There is a new element, the world-sheet $U(1)$ gauge field $A$,
but this can be accomodated as follows.
We work with the
$(0,4)$ supersymmetry transformation
\eqn\gaugesusy{\eqalign{
\delta A_{++} &=i\eta_{+AA'}\psi^{AA'}_+ \cr
\delta A_{--} &=0 \cr
\delta\psi^{AA'}_+ &= \eta_+^{AA'} F
}}
with $F=\p_{--}A_{++}-\p_{++}A_{--}$.
This commutes to a translation up to gauge transformation,
\eqn\comsusy{\eqalign{
[\delta_{\eta'},\delta_{\eta}]A_{++} &= -i\eta\eta'
	(\p_{--}A_{++} + \p_{++}\epsilon) \cr
[\delta_{\eta'},\delta_{\eta}]A_{--} &= -i\eta\eta'
	(\p_{--}A_{--} + \p_{--}\epsilon)
}}
with $\epsilon=-A_{--}$.  This gauge transformation will also enter
into the supersymmetry transformation for matter fermions;
for example
\eqn\matgauge{
\delta\chi^{A\bar m}_- = \eta_+^{AA'} (\p_{--}+iA_{--})\phi^{A'\bar m}.
}

The non-trivial verification is the on-shell $\lambda$ supersymmetry.
Varying twice using \leftsusy\ gives (as in \witadhm\ section 2.2)
\eqn\lamsusy{\eqalign{
[\delta_{\eta'},\delta_{\eta}]\lambda^{\bar M}_+ &=
	-i\eta\eta' G^{\bar M}_\theta \rho^\theta \cr
&= -i\eta\eta'(\p_{--}+iA_{--})\lambda^{\bar M}_+
}}
by the equations of motion.  But this is the combined supersymmetry
and gauge transformation \matgauge.

We then derive the $0$-brane theory by dimensional reduction.
The only role of the $U(1)$ gauge field will be to
restrict the allowed states of the $0$-brane
quantum mechanics to $U(1)$ neutral states such as the adjoint.
It will not play a role in the construction of the self-dual connection,
which proceeds in a way similar to the $SO(N)$ case.

The gauge fermions $\lambda$ have the following Yukawa couplings to $\chi$:
\eqn\IIyuk{
\left(\matrix{\bar\chi_{1}&\bar\chi_{2}}\right)_-.
\left(\matrix{X^{11}-b^{11}&X^{12}-b^{12}&h_{M}^{1\bar m}\cr
	X^{21}-b^{21} &X^{22}-b^{22} &(h^{2~tr})_{\bar m}^{M}}\right)
\left(\matrix{\chi^{1}\cr\chi^{2}\cr\lambda}\right)_+
}
defining an operator $\CD^+$, such that the gauge bundle will be the
kernel $\ker\CD^+$,
spanned by the massless modes $\CD^+v_i=0$.
The complex conjugate fermions have similar Yukawa couplings involving
the adjoint operator,
$\bar\lambda\CD\chi$, and in terms of $\CD$ the gauge bundle will be
the cokernel $\cok \CD$.
The two definitions of subbundle agree, and the final Lagrangian can again
be written in the form \gaugeact.

\newsec{Conclusions}

In the present work, we explicitly verified the claim of Witten that
instanton backgrounds in type \I\ string theory are
equivalent to configurations of Dirichlet $5$-branes.

{}From the point of view of a type \I\ fundamental string, an instanton
background and a configuration of Dirichlet $5$-branes act as two
rather different-looking boundary conformal field theories;
one with the action \direct, the other with an action containing the
DD and DN vertex operators for $X$ and $h$.
Both must renormalize to the same boundary
theory -- surely true on grounds of supersymmetry, but not easy to verify
explicitly.

However, the effect of the D-brane configuration on the Dirichlet $1$-brane
is much simpler.
Its world-sheet theory includes an ``ADHM sigma model''
sector, which reproduces the vector potential of the
equivalent instanton background.

Combining this result with \douglas\ provides an answer to a question
raised by Witten in \witadhm: from the point of view of a string
moving in an instanton background, the existence of a second branch
of minima of the superpotential in the zero scale size limit,
$b=0$ and $\phi\ne 0$, is rather mysterious.
On this branch the string is equivalent to an instanton in the $5$-brane
world-volume gauge theory.  Of course, the metric on this
moduli space is
quantum corrected and it has been argued that the zero scale size limit
is at infinite distance in the quantum moduli space \witstr.

The construction is clearly very general.  For example,
we believe it will be straightforward
to apply it to the construction of $U(N)$ instanton moduli spaces on ALE
spaces in \dm, and reproduce Kronheimer and Nakajima's explicit
construction of gauge fields and metrics \refs{\kn,\etal}.

In fact, the construction is more general than the ADHM construction,
as it does not require \hk structure on the moduli space, but only the
ability to compute the superpotential of the resulting gauge theory.
For example, placing $5$-branes in various planes in six dimensions,
say the $012345$, $012367$ and $012389$ planes, will produce
configurations equivalent to six-dimensional gauge fields preserving a
supersymmetry, i.e. which solve the Donaldson-Uhlenbeck-Yau equation.
The corresponding $1$-brane sigma model will
then provide a construction of the six-dimensional
gauge connection.
\footnote*{
Because this theory has only $(0,2)$ world-sheet
supersymmetry, there might be world-sheet renormalization, and only
in the IR limit are we guaranteed to get a solution of the Donaldson-Uhlenbeck-Yau equation.
It will be interesting to see if this is required, and if so, give a
physical interpretation to this RG flow.
We thank E. Witten for bringing up this point.}

\subsec{Dirichlet branes as local probes}

A more general lesson to draw from this work is that the
Dirichlet brane can serve as a local probe,
which can measure the fields at a point in space-time.
Traditionally, a question such as ``what is the metric (or gauge field)
at the point $x$'' has been regarded as meaningless in string theory,
because a fundamental string provides a probe which is necessarily smeared
out over distances $\sqrt{\alpha'}$.
In calculations, this shows up in the fact that only on-shell vertex
operators give uniquely defined amplitudes.

In string
perturbation theory, a D-brane is effectively an infinitely massive
object, whose center of mass
motion depends on the fields (and perhaps finitely
many derivatives) at a point.
Thus its world-volume theory must -- at least implicitly --
encode this data.

Note that we are not claiming that the world-volume theory does not
receive $\alpha'$ corrections -- rather, that the result of adding such
corrections can be usefully described as a corrected local gauge field
$A_\mu(x)$.

Can one define local fields by taking the limit $\lambda\rightarrow 0$
of finite coupling results?
This seems plausible for gauge fields,
but is not as obvious for the metric.
For example, one might worry that an infinitely massive probe would
necessarily affect the original background.  This question deserves
a detailed analysis (it is non-trivial in field theory as well).

We believe that the use of D-branes as probes will help clarify the
geometric interpretation of other D-brane solutions as well.

\medskip
We would like to thank G. Moore for some
collaboration on this project, 
and M. Berkooz, R. Leigh, N. Nekrasov, S. Shatashvili and E. Witten
for discussions.

\bigskip

\listrefs
\end